\documentclass[aps,prd,preprintnumbers,showpacs,showkeys,nofootinbib,
superscriptaddress,fleqn,floatfix,tightenlines,10pt]{revtex4-1}
\usepackage{amsmath,amsfonts,amssymb,amscd,amsxtra,amsthm}
\usepackage{graphicx}
\usepackage{epstopdf}
\usepackage{bm}
\usepackage{hyperref}
\usepackage[normalem]{ulem} 
\usepackage[dvipsnames]{xcolor} 
\renewcommand\sout{\bgroup \color{red} \ULdepth=-.5ex \ULset}

\usepackage[section]{placeins}
\usepackage[caption = false]{subfig}
\usepackage{multirow}
\newcommand{\be}{\begin{eqnarray}}
\newcommand{\ee}{\end{eqnarray}}

\newcommand{\ket}{\rangle}
\newcommand{\bra}{\langle}

\begin{document}
\preprint{INHA-NTG-07/2019}
\title{Heavy baryon production with an instanton interaction}
\author{Sang-In Shim}
\email[E-mail: ]{shimsang@rcnp.osaka-u.ac.jp}
\affiliation{Research Center for Nuclear Physics (RCNP),
Osaka University, Ibaraki, Osaka, 567-0047, Japan}
\author{Atsushi Hosaka}
\email[E-mail: ]{hosaka@rcnp.osaka-u.ac.jp}
\affiliation{Research Center for Nuclear Physics (RCNP),
Osaka University, Ibaraki, Osaka, 567-0047, Japan}
\author{Hyun-Chul Kim}
\email[E-mail: ]{hchkim@inha.ac.kr}
\affiliation{Department of Physics, Inha University, Incheon 22212,
Republic of Korea}
\affiliation{School of Physics, Korea Institute for Advanced Study 
  (KIAS),\\ Seoul 02455, Republic of Korea}
\date{\today}
\begin{abstract}
We propose a new reaction mechanism for the study of strange and
charmed baryon productions. In this mechanism we consider the
correlation of two quarks in baryons, so it can be called the
two-quark process. As in the previously studied one-quark process, we
find large production rates for charmed baryons in comparison with
strange baryons.  Moreover, the new mechanism causes the excitation of
both the $\rho$ mode and the $\lambda$ mode. 
Using the wave functions for baryons from a quark model, 
we compute the production rates of various baryon states.
We find that the production rates reflect the structure of the wave
functions that imply the usefulness of the reactions for the study of
baryon structures.  
\end{abstract}
\maketitle
\section{Introduction}
Much part of the recent activities in hadron spectroscopy is devoted
to the study of hadrons containing heavy quarks~\cite{Hosaka:2016pey}
(and references therein).  This is largely motivated by a series of
observations of new heavy hadrons~\cite{Choi:2003ue, Acosta:2003zx,
  Aubert:2004ns, Abazov:2008qm, Chatrchyan:2012ni, Aaij:2012da,
  Ablikim:2013mio, Aaij:2015tga, Aaij:2016phn, Aaij:2016ymb,
  Aaij:2016iza, Aaij:2017nav, Yelton:2017qxg, Aaij:2019vzc}, which
have not been expected in the conventional naive quark
model~\cite{GellMann:1964nj,Zweig:1964jf}. 
In order to understand the production mechanism of these newly found
heavy hadrons including the exotic ones, we need to consider more
sophisticated quark-gluon dynamics inside a heavy hadron. 

\begin{figure}[htb]
\includegraphics[width=4cm]{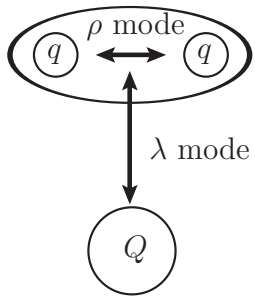}
\caption{A schematic picture of the $\lambda$ and
  $\rho$ modes. Here, two light quarks denoted by $q$'s form a diquark
  and $Q$ stands for a heavy quark} 
\label{fig:1}
\end{figure}

However, one clear virtue of the heavy-light quark systems is the
presence of the heavy quarks. Since the heavy quark has a very large
mass, the kinetic energies of the heavy quarks inside a heavy hadron
are suppressed by the inverse of the heavy-quark mass, which makes 
the quark dynamics inside a heavy baryon simpler than that inside a
light baryon. For example, in a conventional heavy baryon, two light
quarks govern dynamics inside it and can be viewed as a diquark.  On
the other hand, the heavy quark can be regarded as an almost static
color source and makes easily the structure of the heavy baryon
decompose into the two excitation modes, namely, the so-called
$\lambda$ and $\rho$ modes.
As shown in Fig. \ref{fig:1}, the former mode describes the motion of
the light diquark with respect to the heavy quark, and the latter
explains relative motion between the two light quarks. 

The essential features of these modes were discussed long time
ago~\cite{Copley:1979wj} but the experimental data were then not
enough to examine the idea quantitatively. As modern accelerators and
detectors have been developed to perform the experiments with
unprecedented precision, it is interesting to describe the
production of heavy hadrons, based on these two modes.  
Moreove, since the E50 experiment at the J-PARC will soon
measure the charmed baryon productions in the reaction $\pi^- + p \to
D^* + Y_c$ and will yield important information on the structure of
various charmed baryons $Y_c$~\cite{e50}, it is of great importance to
study theoretically the heavy-hadron reactions with these two
different modes considered. Motivated by these discussions, 
we have started the study of the above production
reactions~\cite{Kim:2014qha,Kim:2016cxr, Kim:2016imp}. 

In the present work, we propose a new microscopic mechanism of
hadronic production reactions and investigate how this new mechanism
allows one to understand the baryon structures for the strangeness and
charm productions. 

Though the mass of the strange quark is much
smaller than that of the charm quark, one can consider it effectively
as a \textit{heavy} object in some cases (but not
always)~\cite{Copley:1979wj, Yoshida:2015tia}. In
Ref.~\cite{Copley:1979wj} the mass inversion of $\Lambda(1830)$ and
$\Sigma(1775)$ was used to indicate that the strange quarks are
heavier than the $u$ and $d$ quarks. In Ref.~\cite{Yoshida:2015tia},
in some cases it was shown that the mixing of the $\lambda$-$\rho$
modes is rather small even at the strange quark mass, which indicates
that the strange quark is often effectively considered to be heavy. In
a slightly different context it is also useful to know the cases of
the Skyrme models where the bound-state approaches describe the
properties of the SU(3) hyperons and heavy baryons successfully, the
strange quark being regarded as an 
heavy object~\cite{Callan:1985hy, Callan:1987xt, Ezoe:2016mkp,
  Ezoe:2017dnp} (see also a review~\cite{Weigel:2008zz}). In this 
respect, we can still apply the method of the two modes to both the
single-strange hyperons and singly heavy baryons. 
It is also useful to consider the strangeness sector, because 
strange hadrons can be produced at the J-PARC together charmed
hadrons.  

In this work, we develop a two-quark microscopic process of the baryon
productions: two constituent quarks in a baryon are internally
involved in a production reaction of mesons and baryons by pion beams,
in addition to the one-quark process that was already studied in a 
previous work~\cite{Kim:2014qha,Kim:2016cxr, Kim:2016imp}. This new
mechanism has a virture that one can look into the reaction mechanisms
in a microscopic way. Note that one-quark and two-quark processes are  
similar to one-step and two-step processes, which are often considered
in calculations of nuclear reactions. For example, when a deuteron or
a helium target is scattered off by mesons or photons and then it is
broken into new baryons, one has to take into account both the
one-step and two-step
processes~\cite{YamagataSekihara:2012yv}. Similarly, when the 
charmed hadrons are produced, the large-momentum transfer is
inevitable, which indicates that both the one-quark and two-quark
processes will contribute to the production of charmed hadrons. In
particular, the two-quark process makes it possible to excite both
$\lambda$ and  $\rho$ modes while it is possible to excite only
$\lambda$ modes in one-quark process. 

To formulate and compute reaction matrix elements, we employ a
nonrelativistic quark model (from now on we refer it simply the
quark model) for baryon wave functions and a simple
interaction which involves three quarks, one anti-quark in the
projectile pion and two constituent quarks in the target proton. 
The baryon wave functions are constructed in the heavy-quark basis,
where the total baryon spin is formed by those of light degrees of
freedom (brown muck) and the heavy quark~\cite{Nagahiro:2016nsx}.  
In this way we can see clearly relations between baryon
structures and production rates. This is indeed the main purpose of
the present study. In contrast, to our best knowledge, 
the interaction that can be suitably used for charm or strangeness
productions is not known. Therefore, we shall tentatively employ a
three-quark interaction that is inspired by  't~Hooft for three
flavors~\cite{tHooft:1976}. This is an effective interaction induced
by instanton dynamics~\cite{Diakonov:1985eg,Schafer:1996wv,
  Diakonov:2002fq}, and has been applied to 
the study of meson properties, for instance,~\cite{NamKim,
  Shim:2017wcq, Shim:2018rwv} and baryon
spectrum~\cite{Takeuchi:1990qj, Takeuchi, Takeuchi:2000wn,
  Takeuchi:2000vi}, and heavy
hadrons~\cite{Chernyshev:1994-1995,Musakhanov:2017gym}. 
The instanton-induced interactions were also used phenomenologically
in the description of the proton and antiproton
annihilation~\cite{Polyakov:1994ir}.  In
the present study we employ that interaction 
for $u, d, s$ and for $u, d, c$ quarks. Though its applicability to
production reactions in all details is not clear, we argue that the
most important formula that we will derive in
Eq.~\eqref{eq:MasterFormula} shares common features of the two-quark
process. 

This paper is organized as follows.  
In Section 2, we briefly introduce the general formalism of how one
can introduce the 't~Hooft-like interaction to describe
microscopically the strange and charmed baryon productions. Then we
derive a general formula for the two-quark process for the
productions. In Section 3, we perform numerical calculations and show
the results for forward-angle scattering. We will then discuss
essential features of the production mechanism of the strange and
charmed baryon productions. More general discussions related to
observables such as the angular dependence of the cross sections will
appear elsewhere. The final section is devoted to summary and
conclusions. 

\section{ Formalism }
\begin{figure}[htp]
\includegraphics[width=7cm]{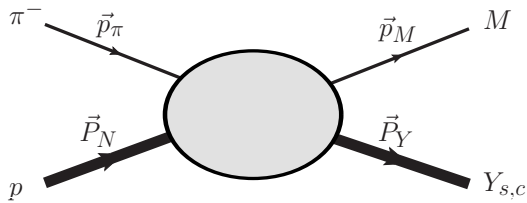}
\caption{Heavy baryon productions from $p\pi^-$ scattering}
\label{fig:2}
\end{figure}

Let us consider the reaction $\pi^- p \to M Y_{s, c}$ as shown  
in Fig.~\ref{fig:2}, where $M$ denotes a $K^0$ or $D^-$ meson with an
anti-strange quark or an anti-charm quark  and $Y_{s, c}$ represents a
heavy baryon with a strange or charm quark. Various kinematic
variables are defined in Fig.~\ref{fig:2}. $\vec{p}_\pi$, $\vec{p}_M$,
$\vec{P}_N$, and $\vec{P}_Y$ stand respectively for the momenta of the
$\pi^-$, the proton ($p$), the meson, and the baryon. 

\begin{figure}[htp]
\captionsetup[subfigure]{labelformat=empty}
\subfloat[]{\includegraphics[width = 3.1in]{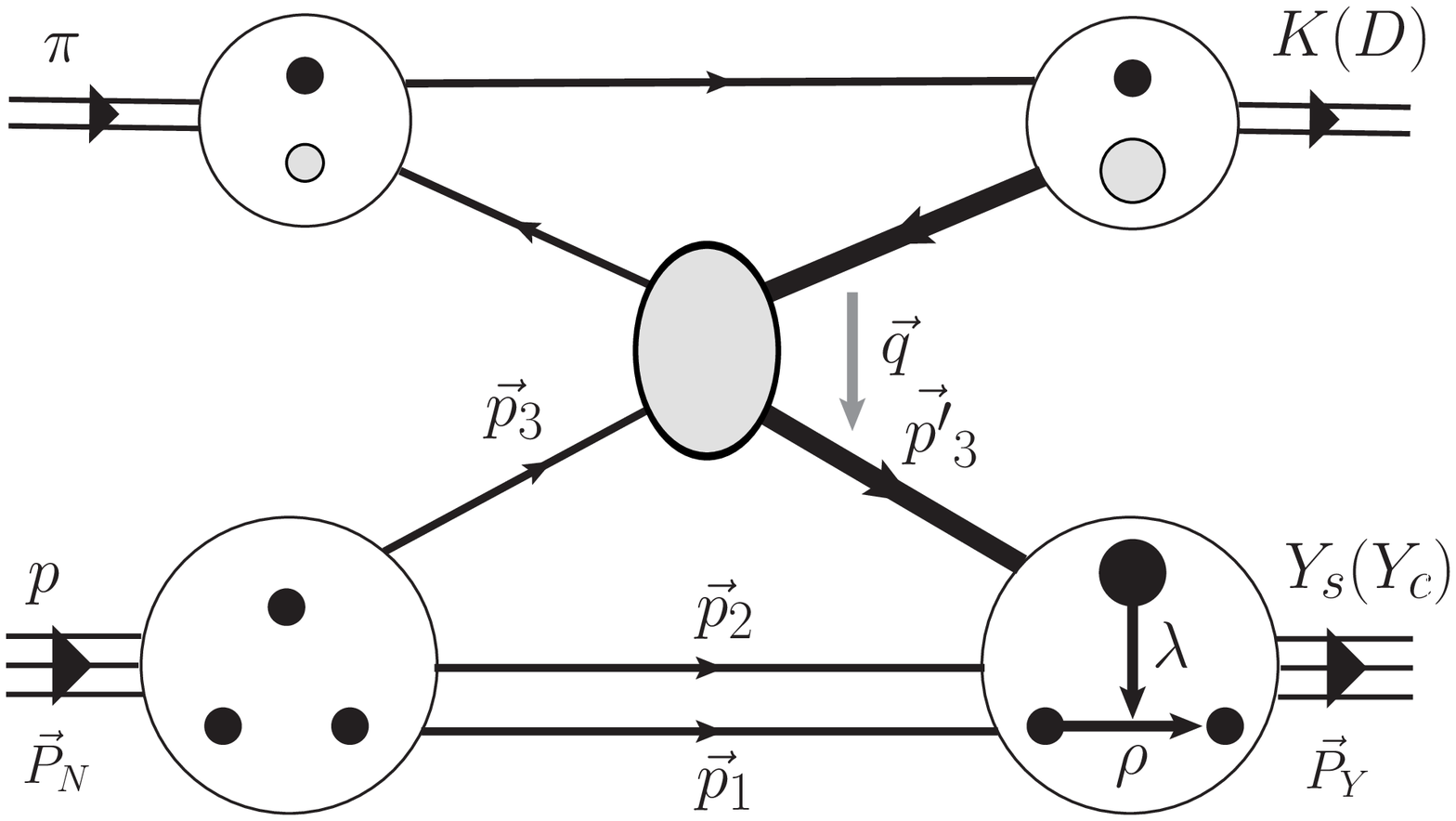}}\hspace{0.1cm}
\subfloat[]{\includegraphics[width = 3.1in]{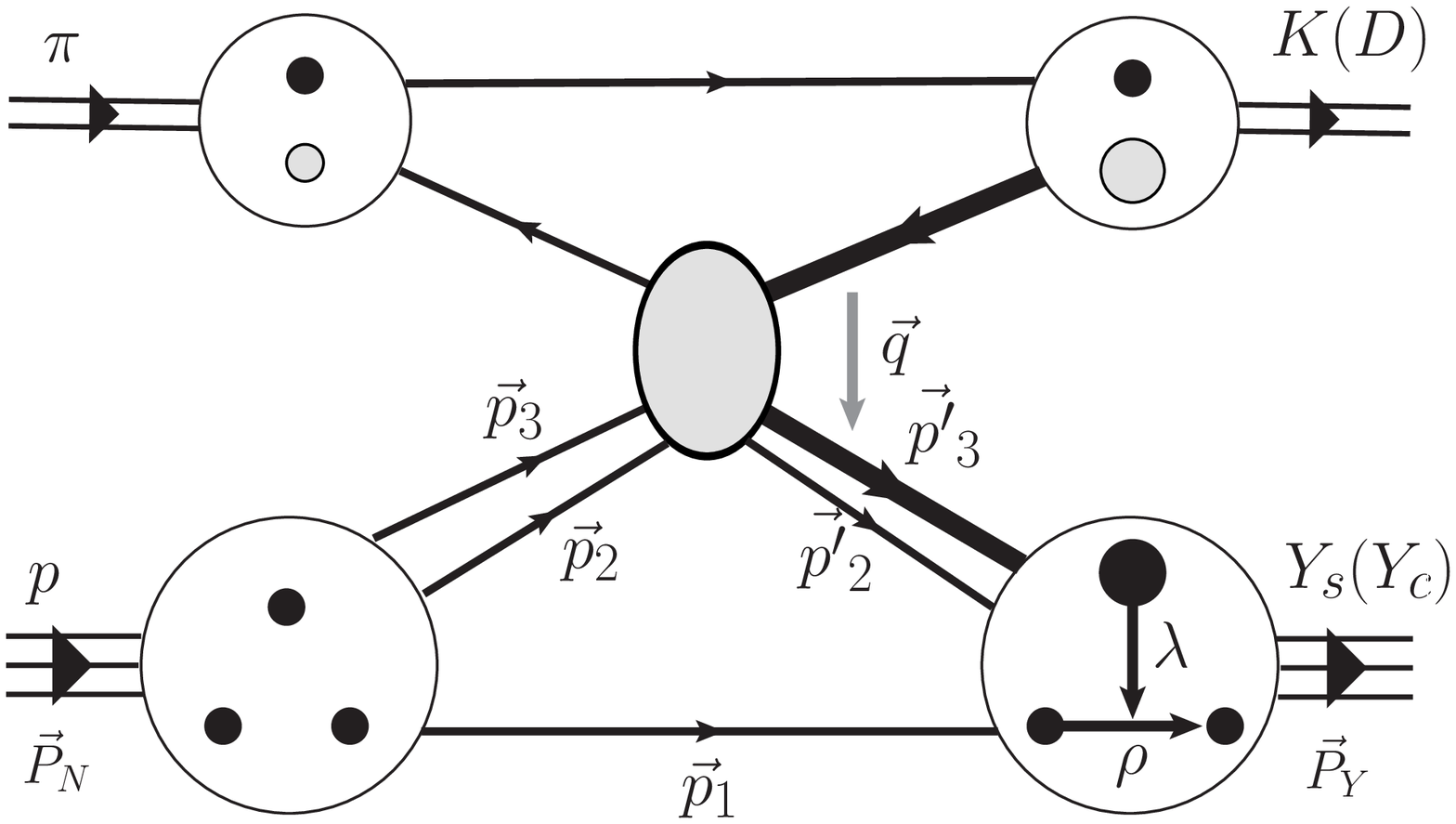}}
\caption{One-quark and two-quark processes for heavy baryon productions. 
Quark-line representations for one-quark (left) and two-quark (right)
processes. The thin lines between inital and final particles represent
light quarks and the thick lines correspond to the heavy
quarks. $\vec{P}_N$ and $\vec{P}_Y$ denote the momenta of the initial
proton and the final state heavy baryons. The momentum $\vec{q}$
stands for the transferred momentum from the inital pion to the heavy
baryon. The momenta $\vec{p}_i$ and $\vec{p'}_i$ ($i = 1,2,3$)
designate the quark momenta inside of the initial and the final states
baryon, respectively.} 
\label{fig:3}
\end{figure}

In Fig.~\ref{fig:3}, we draw the quark-line representations for one-quark
and two-quark processes on the left and right panels, respectively. 
In the one-quark process, an antiquark in the pion annihilates with
one quark in the proton, and an $s\bar s$ or $c \bar c$ pair is
created, while in the two-quark process, an antiquark in the pion
interacts with two-quarks in the proton.  From these pictures, we see
that one-quark process excites only $\lambda$ modes, while the
two-quark process excites both $\lambda$ and $\rho$ modes. 

In Fig.~\ref{fig:3}, we also show momentum fractions carried by
various quarks: the momenta of the initial and the final state baryons
consist of the momenta of the three quarks inside the baryons,
$\vec{P}_N=\vec{p}_1+\vec{p}_2+\vec{p}_3$,
$\vec{P}_Y=\vec{p}_1+\vec{p'}_2+\vec{p'}_3$, where $\vec{p}_i$ and
$\vec{p'}_i=\vec{p}_i+\vec{q}_i$ ($i=1,2,3$) are the quark momenta
inside of the baryons and $\vec{q}_i$ is the transferred momentum from
the initial pion to the $i$-th quark in the heavy baryon.  In the two
quark process the momentum transfer $\vec{q}$ is shared by two quarks
(2, 3), so that $\vec{q} = \vec{P}_Y-\vec{P}_p = \vec{q}_2 +
\vec{q}_3$ becomes the transferred momentum from the pion to the heavy 
baryon. Since the one-quark process has been studied
previously~\cite{Kim:2014qha,Kim:2016cxr,Kim:2016imp}, we will focus
on the two-quark process in the following subsections and the next
sections.    
\subsection{Three-quark interaction}
In this subsection, we discuss briefly several features of the
't~Hooft-like interaction, which will be useful for discussions of
various production rates. An advantage of this interaction is that the
reaction occurs at one place (single step),  
which makes the computation of matrix elements easy. 
The 't~Hooft-like interaction is for three quarks with three flavors,
$N_f=3$, which arises from the instanton
dynamics of QCD~\cite{tHooft:1976, Diakonov:1985eg, Schafer:1996wv,
  Diakonov:2002fq}. 
In general, it is a nonlocal interaction in which
the dynamical quark mass is momentum-dependent. Moreover, the $2N_f$
quark-quark interaction considers only the light flavors, i.e. the up,
down, and strange quarks. When one includes heavy quarks together with
the light quarks, one has to derive the heavy-light quark interactions
from the instanton vacuum again. Though there are some theoretical
works on this heavy-light quark interactions from the instanton
vacuum~\cite{Chernyshev:1994-1995, Musakhanov:2017gym}, its
applicability is not sufficiently matured. Thus, in the present work, 
we will consider a simplified version of the 't Hooft-like interaction
including strange or charm quarks. Actually, it is also possible to
transform this simplified one into a form of the heavy-light quark
interaction similar to that of~\cite{Chernyshev:1994-1995,
  Musakhanov:2017gym}.  We will also take a local 
form of the 't Hooft-like interaction.

We start from the 't Hooft-like six-quark interaction defined
by~\cite{tHooft:1976}
\begin{align}
\mathcal{L}_{tH}
= c \,\mathrm{det}[\bar{q}_i (1+\gamma_5)q_j] + H.c. 
= c \left| \begin{array}{ccc}
\bar{u}(1+\gamma_5)u & \bar{u}(1+\gamma_5)d & \bar{u}(1+\gamma_5)s \\
\bar{d}(1+\gamma_5)u & \bar{d}(1+\gamma_5)d & \bar{d}(1+\gamma_5)s \\
\bar{s}(1+\gamma_5)u & \bar{s}(1+\gamma_5)d & \bar{s}(1+\gamma_5)s
 \end{array} \right| + H.c.,
 \label{eq:tHooft}
\end{align}
where $c$ is an interaction strength. In general,it is difficult to
predict the absolute magnitudes of the reaction cross
sections. Therefore, we treat the strength $c$ as a free 
parameter. On the other hand, we can discuss at least the ratios of
various cross sections rather than their absolute values, so we will
focus on the ratios in the present paper. 

It is convenient to rewrite Eq.~\eqref{eq:tHooft} by using the
Fierz transformation to rearrange six quarks by observing the
followings: The $\bar u$ field annihilates the $\bar u$ state in the
incoming $\pi^-$, the $s$ field creates the $\bar s$ state in the
produced $K$ meson, the $d$ fields annihilates the corresponding
quarks in the proton, and the $\bar d$ and $\bar s$ fields create the
corresponding ones in the strange baryon.  Thus, the 't Hooft-like
interaction can be reexpressed as 
\begin{align}
\mathcal{L}_{tH}
&=  4\,\mathrm{det}[\bar{q}_{i L}q_{jR}] + H.c. \nonumber \\
&= 4 \Bigg[ 
\bigg(1+\frac{1}{N_c}\bigg) 
(\bar{u}_L s_R)
\Big( (\bar{d}_L u_R)(\bar{s}_L d_R)
-(\bar{d}_L d_R)(\bar{s}_L u_R)\Big)\nonumber \\
&+\frac{1}{8 N_c}(\bar{u}_L \sigma^{\mu \nu} s_R) 
\Big( (\bar{d}_L \sigma_{\mu\nu} u_R)(\bar{s}_L d_R)
+(\bar{d}_L u_R)(\bar{s}_L \sigma_{\mu\nu} d_R)
-(\bar{d}_L \sigma_{\mu\nu} d_R)(\bar{s}_L u_R)
-(\bar{d}_L d_R)(\bar{s}_L \sigma_{\mu\nu} u_R)\Big) \cr
&
+(\bar{u}_L \frac{\lambda^i}{2} s_R)
\Big(
(\bar{d}_L \frac{\lambda^i}{2} u_R)
(\bar{s}_L d_R)
+
(\bar{d}_L u_R)
(\bar{s}_L \frac{\lambda^i}{2} d_R)
- (\bar{d}_L \frac{\lambda^i}{2} d_R)
(\bar{s}_L u_R)
-
(\bar{d}_L d_R)
(\bar{s}_L \frac{\lambda^i}{2} u_R)
\Big)\nonumber \\
&
+\frac{1}{4}(\bar{u}_L \sigma^{\mu\nu}\frac{\lambda^i}{2} s_R) 
\Big(
(\bar{d}_L \sigma_{\mu\nu}\frac{\lambda^i}{2} u_R)
(\bar{s}_L d_R)
+
(\bar{d}_L u_R)
(\bar{s}_L \sigma_{\mu\nu}\frac{\lambda^i}{2} d_R)
                    \cr
& \hspace{1cm}-
(\bar{d}_L \sigma_{\mu\nu}\frac{\lambda^i}{2} d_R)
(\bar{s}_L u_R)
-
(\bar{d}_L d_R)
(\bar{s}_L \sigma_{\mu\nu}\frac{\lambda^i}{2} u_R)
\Big)\Bigg]\cr
&\hspace{0.5cm} + \mbox{H.c.},
\label{eq:tHooftexp}
\end{align}
where $\bar{q}_{iL}$ and $q_{jR}$ denote the left- and right-handed quark
fields, $q_{iR}=(1+\gamma_5)q_i/2$ and $\bar{q}_{iL} = \bar{q}_{j}(1 +
\gamma)/2$ and $\lambda^i$ are the SU(3) Gell-Mann matrices defined in
color space.  Since the mesons and baryons in the initial and the
final states should be color singlets, the terms with $\lambda^i$ in
Eq.~\eqref{eq:tHooftexp} do not contribute to the present reaction.  
Considering suitable leading-oder terms in the $1/N_c$ expansion, we 
need only the following terms 
\begin{align}
\mathcal{L}_{tH}
\rightarrow& \, 4 \,
(\bar{u}_L s_R)
\Big( (\bar{d}_L u_R)(\bar{s}_L d_R)
-(\bar{d}_L d_R)(\bar{s}_L u_R)\Big)\nonumber + H.c. \nonumber \\
=&\,
(\bar{u}s)
\Big[\,
(\bar{d}u\big)(\bar{s}d) + (\bar{d}\gamma_5 u)(\bar{s}\gamma_5 d)
- (\bar{d}d\big)(\bar{s}u) - (\bar{d}\gamma_5 d)(\bar{s}\gamma_5 u) 
\, \Big]  \nonumber \\
&+
(\bar{u}\gamma_5 s) 
\Big[\,
 (\bar{d}u\big)(\bar{s}\gamma_5 d) + (\bar{d}\gamma_5 u)(\bar{s}d)
- (\bar{d}d\big)(\bar{s}\gamma_5 u) - (\bar{d}\gamma_5 d)(\bar{s}u) 
\, \Big] .
\label{eq:tHooftLO}
\end{align}
In this expression, only the terms in the second line are
relevant, because the meson matrix elements of $(\bar u s)$ in the
first line vanish in the production reaction of a pseudoscalar meson
due to parity conservation. For baryon matrix elements in the 
nonrelativistic quark model, we need expressions in terms of two
component spinors. We have explicitly computed the $(\bar u s)$ term
of Eq.~\eqref{eq:tHooftLO} and found that the relevant operators
are reduced to the identity operators. This can be verified by
neglecting the Fermi motion of the quarks confined in baryons and for
forward-angle scattering which is the dominant component of the
reactions that we study in this paper. Therefore the operator that we
need is written as 
\begin{align}
\mathcal{L}_{tH}
\rightarrow&\,
(\bar{u}s)
\Big[\,
(d^{\dagger}u\big)(s^{\dagger}d) 
- (d^{\dagger}d\big)(s^{\dagger}u)
\, \Big]
\equiv \,
\mathcal{O}^{M}\cdot\mathcal{O}^{B},
\label{eq:tHooftLO2}
\end{align}
where $\mathcal{O}^{M} \sim \bar u s$ acts on the meson transition, 
$\pi \to K$, whereas $\mathcal{O}^{B}$ on the baryon transition, $p
\to Y$, and $u$, $d$, $d^{\dagger}$, $s^{\dagger}$ are two
component spinors for the quarks in a baryon. 

\subsection{Baryon wave functions}
As mentioned previously, we employ the baryon wave functions taken
from the quark model. In the limit of infinitely heavy-quark mass
($m_Q\rightarrow \infty$), the spin of the heavy quark $s_Q$ is conserved,
which leads to the conservation of the light-quark spin $j$. It is
known as the heavy-quark spin symmetry. Thus, we construct the baryon
wave functions that are the simultaneous eigenstates of $j$ and $s_Q$
to describe the baryon with one heavy (strange or charm) quark (for
more explanation, we refer to Refs.~\cite{Nagahiro:2016nsx,
  Yoshida:2015tia}).  In the quark model, a baryon wave function is given as
a product of the orbital, spin, flavor and color parts as follows: 
\begin{align}
|\Psi \rangle 
= |\mathrm{orbit} \rangle \otimes |\mathrm{spin}\rangle \otimes |
  \mathrm{flavor} \rangle \otimes | \mathrm{color}\rangle.
\end{align}
Since the color part is always antisymmetric, the rest of the
baryon wavefunction should be taken to be totally symmetric. 
Note that the interaction Lagrangian in Eq.~\eqref{eq:tHooftLO2} is
given as a color singlet and a scalar in spin space.

Introducing the quark potential of the harmonic-oscillator type for
confinement, we can decompose the orbital wavefunction into 
those of the center-of-mass (CM) $\vec{X}$ and of internal coordinates
$\vec{\lambda}$, $\vec{\rho}$ as  
\begin{align}
&\Psi_N(\vec{x}_1,\vec{x}_2,\vec{x}_3) 
= e^{i\vec{P}_N \cdot
                \vec{X}}\psi^\rho_0(\vec{\rho})\psi^\lambda_0(\vec{\lambda}),
                \nonumber \\ 
&\Psi_Y(\vec{x}_1,\vec{x}_2,\vec{x}_3) 
= e^{i\vec{P}_Y \cdot \vec{X}}\psi^\rho_{n_\rho l_\rho m_\rho}(\vec{\rho})
 \psi^\lambda_{n_\lambda l_\lambda m_\lambda}(\vec{\lambda}), 
\label{eq:BaryonWF}
\end{align}
where $\vec{X}$, $\vec{\rho}$, $\vec{\lambda}$ are related to
$\vec{x}_1$, $\vec{x}_2$,  $\vec{x}_3$, respectively, as
\begin{align}
\vec{X} &=\frac{1}{2m_q + m_Q}\Big( m_q (\vec{x}_1 + \vec{x}_3) + m_Q
          \vec{x}_3\Big), 
\nonumber \\
\vec{\rho} & = \vec{x}_2 - \vec{x}_1
\nonumber \\
\vec{\lambda}&=\frac{1}{2}(\vec{x}_1+\vec{x}_2)-\vec{x}_3.
\end{align}
Here the light quarks are labeled by 1 and 2, and the heavy quark by
3. Assuming isospin symmetry, we can express the quark masses as
$m_1=m_2=m_q<m_3=m_Q$. 
The internal wavefunctions $\psi^\rho_{n_\rho l_\rho
  m_\rho}(\vec{\rho})$ and $\psi^\lambda_{n_\lambda l_\lambda
  m_\lambda}(\vec{\lambda})$ are typically written as 
\begin{align}
\psi_{n l m}(\vec{r}) = R_{n l}(r) Y_{lm}(\hat{r}),
\end{align}
where $Y_{lm}(\hat{r})$ denote the spherical harmonics and
$R_{nl}(\vec{r})$ stand for the radial wavefunctions, which are given
explicitly in Appendix A.  
The wavefunction $\psi_{0}(\vec{r})$ represents the ground state with 
$n=l=m=0$ and $\psi_{n l m}(\vec{r})$ for the ground state and excited
states with quantum numbers $n$, $l$, $m$.  
From now on, $\psi_{n l m}(\vec{r})$ will be written compactly by
$\psi_{l}(\vec{r})$, because we will consider only the excitations of
$l$ in the present work. In a more realistic model containing the
linear confining potential with the spin-spin interaction, the
$\lambda$ and $\rho$ modes are mixed each other.  However, in 
Ref.~\cite{Yoshida:2015tia} it was shown that some baryon states are
dominated by either the $\lambda$ or the $\rho$ mode. Moreover, once
we know the properties of $\lambda$ and $\rho$ modes separately, the
realistic cases of their mixing can be estimated. Because of these
reasons in the present study we consider various matrix elements for
the $\lambda$ and $\rho$ modes separately. 

The flavor (isospin) parts of the heavy baryons will be expressed by
$D^{I}_{I_z}Q$. For $I=0$ 
\begin{align}
D^{0}_{0}Q = \frac{1}{\sqrt{2}}(|ud\rangle-|du\rangle)Q,
\end{align}
and for $I=1$ and $I_z=0$
\begin{align}
D^{1}_{0}Q = \frac{1}{\sqrt{2}}(|ud\rangle+|du\rangle)Q,
\end{align}
where $Q$ stands for a heavy quark.

Similarly, the spin part of the diquark can be expressed by
$d^{s}_{s_z}$, 
\begin{align}
d^{0}_{0} = \frac{1}{\sqrt{2}}(|\uparrow \downarrow\rangle -
  |\downarrow \uparrow\rangle), 
\end{align}
\begin{align}
d^{1}_{s_z} = \left\{\begin{array}{cc}
|\uparrow \uparrow\rangle,  & s_z =1,\\
\frac{1}{\sqrt{2}}(|\uparrow \downarrow\rangle + |\downarrow
                       \uparrow\rangle), & s_z=0, \\ 
|\downarrow \downarrow\rangle, & s_z=-1,
\end{array}\right.
\end{align}
where $s$ designates the spin angular momentum of the diquark and
$s_z$ corresponds to its $z$-th component. The spin part of a heavy
quark is denoted by $\chi_Q$. 

By using these expressions, the baryon wavefunctions of $\Lambda_Q$
and $\Sigma_Q$ with total spin $J$ can be written as 
\begin{align}
|\Lambda_Q(J, J_{z})\rangle & =
 [[\psi^{\rho}_{l_\rho}(\vec{\rho})\psi^{\lambda}_{l_\lambda}
  (\vec{\lambda}),d]^j,  \chi_Q]^J_{J_{z}} D^0 Q
\label{e:WFLambda}
  \\ 
  |\Sigma_Q(J, J_{z})\rangle &= [[\psi^{\rho}_{l_\rho}(\vec{\rho})
\psi^{\lambda}_{l_\lambda}(\vec{\lambda}),d]^j, \chi_Q]^J_{J_{z}} D^1 Q,
\label{e:WFSigma}
\end{align}
where $[l_1,l_2]^{l_3}$ represents angular momentum coupling of
$l_1+l_2=l_3$ with Clebsh-Gordan coefficients included properly, and
the color and CM parts of the wavefunctions are not included.  

The SU(6) proton wavefunction with $J_z=1/2$ is given as
\begin{align}
  |p(1/2,1/2) \rangle = \psi^{\rho}_0(\vec{\rho})
  \psi^{\lambda}_0(\vec{\lambda}) \frac{1}{\sqrt{2}}\Big(
\chi^{\rho}_{1/2}\phi^{\rho} + \chi_{1/2}^{\lambda}\phi^{\lambda}
 \Big)
\end{align}
where the spin and isospin wavefunctions, $\chi^{\rho,
  \lambda}_{1/2}$ and $\phi^{\rho, \lambda}$ are given respectively by 
\begin{align}
\chi^{\rho}_{\frac{1}{2}} 
& = \frac{1}{\sqrt{2}} 
(|\uparrow \downarrow \uparrow \rangle 
- |\downarrow \uparrow \uparrow \rangle), \\
\chi^{\lambda}_{\frac{1}{2}}
& = \frac{-1}{\sqrt{6}}
(|\uparrow \downarrow \uparrow \rangle 
+ |\downarrow \uparrow \uparrow \rangle
- 2 |\uparrow \uparrow \downarrow \rangle),
\end{align}
and
\begin{align}
\phi^{\rho}
& = \frac{1}{\sqrt{2}} 
(|u d u \rangle - |d u u \rangle), \\
\phi^{\lambda}
& = \frac{-1}{\sqrt{6}}
(|u d u \rangle + |d u u \rangle - 2 |u u d \rangle).
 \end{align}

\subsection{Transition amplitudes}
The transition amplitude for the reaction $\pi^- p \to M Y$ is 
written as a factorized form 
\begin{align}
\int d^4 x \bra \, Y \, M | \mathcal{L}_{tH}  \, | \, N \, \pi^- \ket
\sim 
\bra M  | \mathcal{O}^{M}  \, | \pi^-  \ket 
\bra Y  | \mathcal{O}^{B}  \, | N  \ket 
\,
2\pi \, \delta(E_Y + E_M - E_N - E_\pi), 
\end{align}
where the baryon part is only the relevant one in the following
discussion. In the two-quark process, the operator $\mathcal{O}^B$ is
a two-body operator and is written as 
\begin{align}
\sum_{ij}\mathcal{O}^B(i,j)
\end{align}
where $i,j=1,2,3$ denote the quark numbers. 
Fixing the number of the heavy quark as 3, we have only two terms 
\begin{align}
\sum_{ij}\mathcal{O}^B(i,j)\,\longrightarrow \, \mathcal{O}^B
  (1,3)+\mathcal{O}^B(2,3). 
\end{align}
The operator has the flavor dependence as in Eq.~(\ref{eq:tHooftLO2}),
while the spin part becomes trivial because it is a scalar.  
Therefore, the baryon matrix element is given by
\begin{align}
&\bra Y  | \mathcal{O}^{B}  \, | N  \ket 
 \nonumber \\
&=\,  
\int d^3 x_1 \, d^3 x_2 \, d^3 x_3 
\Psi^*_Y(\vec{x}_1,\vec{x}_2,\vec{x}_3) 
_{Y}\langle \mathrm{spin |} {\otimes} _{Y}\langle \mathrm{flavor} |
\Big[ \, \mathcal{O}^{B}(1,3) \, + \mathcal{O}^{B}(2,3) \, \Big]
| \mathrm{spin} \rangle_{N}{\otimes} |\mathrm{flavor} \rangle_{N}
\Psi_N(\vec{x}_1,\vec{x}_2,\vec{x}_3) \nonumber \\
&=
\frac{C_{Y}}{2}
\, 
\int d^3 x_1 \, d^3 x_2 \, d^3 x_3  
\,  \Psi^*_Y(\vec{x}_1,\vec{x}_2,\vec{x}_3) 
\Big[ \, 
 e^{i\vec{q}\cdot\vec{x_1}}
\, \delta^{(3)}(\vec{x}_1-\vec{x}_3)
+ (1 \leftrightarrow 2)
\, \Big]
\Psi_N(\vec{x}_1,\vec{x}_2,\vec{x}_3).
\label{eq:Amplitude}
\end{align}
Note that we have carried out the calculation in the coordinate space
of three quarks $x_1, x_2, x_3$.  The two-quark operator 
$\mathcal{O}(i,j)$ acts on the $i$-th and $j$-th quarks.  In the
second equality, the delta function indicates that the interaction occurs
at a single point. The spin-isospin factor $C_Y$ arises from the
Clebsch-Gordan coefficients in the computations of spin and flavor
matrix elements. The factor 1/2 was introduced for convenience.  

Using the identity
\begin{align}
e^{i\vec{q}\cdot\vec{x_1}}
\, \delta^{(3)}(\vec{x}_1-\vec{x}_3)
= \int d^3q_1\,  d^3q_3\,  
e^{i\vec{q_1} \cdot\vec{x_1}} e^{i\vec{q_3} \cdot\vec{x_3}}
\delta^{(3)}\big(\vec{q} -\vec{q}_1-\vec{q}_3\big), 
\end{align}
one can rewrite the transition amplitude as
\begin{align}
&\bra Y  | \mathcal{O}^{B}  \, | N  \ket \nonumber \\
&=\frac{C_Y}{2}
\int d^3 q_1\,d^3 q_3 
\,\delta^{(3)}\big(\vec{q} -\vec{q}_1-\vec{q}_3\big)
\int d^3 x_1 \,d^3 x_2 \,d^3 x_3  
\, \Psi^*_Y(\vec{x}_1,\vec{x}_2,\vec{x}_3) 
 e^{i\vec{q_1} \cdot\vec{x_1}} e^{i\vec{q_3} \cdot\vec{x_3}}
\Psi_N(\vec{x}_1,\vec{x}_2,\vec{x}_3) 
\nonumber \\
&\hspace{0.3cm}
+\frac{C_Y}{2}
\int d^3 q_2\,d^3 q_3 
\,\delta^{(3)}\big(\vec{q} -\vec{q}_2-\vec{q}_3\big)
\int d^3 x_1 \,d^3 x_2 \,d^3 x_3  
\, \Psi^*_Y(\vec{x}_1,\vec{x}_2,\vec{x}_3) 
 e^{i\vec{q_2} \cdot\vec{x_2}} e^{i\vec{q_3} \cdot\vec{x_3}}
\Psi_N(\vec{x}_1,\vec{x}_2,\vec{x}_3) \nonumber \\
&=\,\delta^{(3)}\big(\vec{P}_Y -\vec{P}_N-\vec{q}\big)
\nonumber \\
&\hspace{0.3cm}\times \frac{C_Y}{2} \int d^3 q_1\,d^3 q_3 
\,\delta^{(3)}\big(\vec{q} -\vec{q}_1-\vec{q}_3\big)
\int d^3 \rho e^{i \vec{q}_{\rho} \cdot \vec{\rho}}
\psi^{\rho *}_{l_{\rho}}(\vec{\rho}) \psi^{\rho }_{0} (\vec{\rho})
\int d^3 \lambda e^{i \vec{q}_{\lambda} \cdot \vec{\lambda}}
\psi^{\lambda' *}_{l_{\lambda}}(\vec{\lambda}) \psi^{\lambda}_{0} (\vec{\lambda})
\nonumber \\
&\hspace{0.2cm}
 + (1 \leftrightarrow 2, \, \vec{\rho} \rightarrow -\vec{\rho}) 
 \label{eq:MasterFormula}
\end{align}
where $\vec{q}_{\rho}=\frac{1}{2}\vec{q}_{1}$ and
$\vec{q}_{\lambda}=\vec{q}_{1}+\vec{q}_{\mathrm{eff}}$ with the
effective momentum transfer defined as 
\begin{align}
  \vec{q}_{\mathrm{eff}} \equiv\frac{m_d }{m_d + m_q}
  \vec{P}_N - \frac{m_d }{m_d + m_Q}\vec{P}_Y.
\end{align}
Having performed the integration over $q_1$ and $q_3$, we obtain the
matrix elements for the productions of the ground-state heavy baryon
as  
\begin{align}
\bra Y(l_\lambda=l_\rho=0) | \mathcal{O}^{B}  \, | N
  \ket  &=C_{Y} \, I_{g.s.} \, (2\pi)^3\delta^{(3)}\big(\vec{P}_Y
          -\vec{P}_p-\vec{q}\big), 
\label{eq:TransAmpGn}
\end{align}
where $I_{g.s.}$ is defined by
\begin{align}
I_{g.s.} &\equiv
\int d^3 \kappa \,
\int d^3 \rho e^{i \frac{1}{2}\vec{\kappa} \cdot \vec{\rho}}
\psi^{\rho *}_{0}(\vec{\rho}) \psi^{\rho }_{0} (\vec{\rho})
\int d^3 \lambda e^{i (\vec{\kappa}+\vec{q}_{\mathrm{eff}}) \cdot \vec{\lambda}}
\psi^{\lambda' *}_{0}(\vec{\lambda}) \psi^{\lambda}_{0} (\vec{\lambda})\nonumber\\
&=
\left(
\frac{16\pi \alpha^2_{\rho} \alpha_{\lambda'}\alpha_{\lambda}}{
	B^2}
\right)^{3/2}
e^{-q^2_{\mathrm{eff}}/(4 B^2)}.
\label{eq:Igs}
\end{align}
Here, $B^2$ is defined by
\begin{align}
B^2 \equiv \frac{8 \alpha_\rho^2+\alpha_{\lambda'}^2+\alpha_{\lambda}^2}{2} 
\end{align}
where $m_d$ denotes the effective mass of a diquark, $\alpha_{\rho}$,
$\alpha_{\lambda}$, and $\alpha_{\lambda'}$ given in Appendix A are
the oscillator parameters for the $\rho$ modes, initial and final
state $\lambda$ modes, respectively. 
Except for the delta function, the matrix elements given in
Eq.~\eqref{eq:TransAmpGn} depend on $\vec{q}_{\mathrm{eff}}$ instead 
of $\vec{q}$ because the recoil effect occurs by the difference
between the masses of particles in initial and final states. In
Eq.~\eqref{eq:Igs},  we have seen that the Gaussian form factor
$\exp\big(-q^2_{\rm{eff}}/(4B^2)\big)$ arises as a consequence of the
use of the harmonic oscillator wave functions. In a realistic
situation, a dipole type $1/\big(1 + q^2_{\rm{eff}}/ (4B^2)\big)$
would be more preferable. Here in our discussions, however, we 
mostly treat the relative strengths of various transitions, where the
form factors are almost canceled out and an actual form of the form 
factor does not affect the conclusion of the present work, as
discussed below.  

For the excited baryons in forward-angle scattering, the matrix elements
are written as 
\begin{align}
\bra Y(l_\lambda=1,l_\rho=0) | \mathcal{O}^{B}  \, | N  \ket 
&=C_{Y} \, I_{l_\lambda = 1} \,
(2\pi)^3\delta^{(3)}\big(\vec{P}_Y -\vec{P}_p-\vec{q}\big)
\label{eq:TransAmpLambda}
\end{align}
and
\begin{align}
\bra Y(l_\lambda=0,l_\rho=1) | \mathcal{O}^{B}  \, | N  \ket 
&=C_{Y} \, I_{l_\rho = 1} \, 
(2\pi)^3\delta^{(3)}\big(\vec{P}_Y -\vec{P}_p-\vec{q}\big)
\label{eq:TransAmpRho}
\end{align}
where $I_{l_{\lambda=1}}$ and $I_{l_{\rho=1}}$ are defined by 
\begin{align}
I_{l_\lambda=1} &\equiv 
\int d^3 \kappa \,
\int d^3 \rho e^{i \frac{1}{2}\vec{\kappa} \cdot \vec{\rho}}
\psi^{\rho *}_{0}(\vec{\rho}) \psi^{\rho }_{0} (\vec{\rho})
\int d^3 \lambda e^{i (\vec{\kappa}+\vec{q}_{\mathrm{eff}}) \cdot \vec{\lambda}}
\psi^{\lambda' *}_{1}(\vec{\lambda}) \psi^{\lambda}_{0} (\vec{\lambda})\nonumber\\
&=
\frac{i\sqrt{2}\alpha_{\lambda'} |\vec{q}_{\mathrm{eff}}|}{
	2 B^2}
\left(
\frac{16\pi \alpha^2_{\rho} \alpha_{\lambda'}\alpha_{\lambda}}{
	B^2}
\right)^{3/2}
e^{-q^2_{\mathrm{eff}}/(4 B^2)},
\label{eq:Ilambda1}
\\
I_{l_\rho=1} &\equiv 
\int d^3 \kappa \,
\int d^3 \rho e^{i \frac{1}{2}\vec{\kappa} \cdot \vec{\rho}}
\psi^{\rho *}_{1}(\vec{\rho}) \psi^{\rho }_{0} (\vec{\rho})
\int d^3 \lambda e^{i (\vec{\kappa}+\vec{q}_{\mathrm{eff}}) \cdot \vec{\lambda}}
\psi^{\lambda' *}_{0}(\vec{\lambda}) \psi^{\lambda}_{0} (\vec{\lambda})\nonumber\\
&=
\frac{-i\sqrt{2}\alpha_{\rho} |\vec{q}_{\mathrm{eff}}|}{
	B^2}
\left(\frac{16\pi \alpha^2_{\rho} \alpha_{\lambda'}\alpha_{\lambda}}{
	B^2}\right)^{3/2}
e^{-q^2_{\mathrm{eff}}/(4 B^2)}.
\label{eq:Irho1}
\end{align}

In order to evaluate the production rates, we also need the meson matrix
elements $\bra M | \mathcal{O}^M| \pi^- \ket$.   This depends also on
the properties of the mesons involved.  However, considering the fact
that the meson states in both the initial and final states are the
same and assuming that the results depend mildly on meson form
factors, we are able to ignore the matrix elements $\bra M |
\mathcal{O}^M| \pi^- \ket$ for the study of relative production rates
of various baryons. Thus, the differential cross sections are computed by
\begin{align}
\mathcal{R}
&=\frac{1}{\mathrm{Flux}} \times |t_{fi}|^2\times \text{Phase space}\\
&\sim\frac{1}{\mathrm{Flux}} \times |C_{Y} I_{l}|^2\times \text{Phase
 space}, 
\end{align}
where $t_{fi}$ denotes the transition amplitudes from the proton state
($i \sim p$) to various heavy-baryon states ($f\sim Y_s$ or $Y_c$). 
In the CM frame, this can be written as
\begin{align}
\mathcal{R}\big(Y(J^{p}, J_z) \big)
\sim\frac{1}{4|p_{i}|\sqrt{s}} |C_{Y}|^2 |I_l|^2
  \frac{|\vec{p}_f|}{4\pi \sqrt{s}}, 
\label{eq:ProdRates}
\end{align}
where $s$ denotes the Mandelstam variable $s=\big(\vec{p}_\pi +
\vec{P}_N\big)^2 = \big(\vec{p}_M + \vec{P}_Y\big)^2$. 

We note that the main formula that we have derived, from
Eq.~\eqref{eq:MasterFormula} to Eq.~\eqref{eq:Irho1},
are for the ’t Hoot-liked interaction which is unity in spin space in
the non-relativistic approximation, namely, $\mathcal{O}^B\sim 1$ in
Eq.~\eqref{eq:Amplitude}.  
These formulae still hold for other types of the interactions with the
operator $\mathcal{O}^B$ suitably changed. If it has spin dependence,
its effect is included in the spin-isospin factor $C_Y$.   
\section{Results and discussions}
\subsection{Kinematic conditions}
We are now in a position to present the numerical results and discuss them.
Since this is the first work on the two-quark process in the
heavy-baryon productions, we will consider only the case of
forward-angle scattering for simplicity. The angular dependence and
other observables will be studied in future works. To demonstrate the 
production rates, we first fix the momentum of the pion at
$k^{\mathrm{Lab}}_\pi = 5\, \mathrm{GeV}$ for strange baryons and
$k^{\mathrm{Lab}}_\pi = 20\, \mathrm{GeV}$ for charmed baryons. These
values of the momenta will provide already sufficient energies to
create the $s\bar s$ or $c \bar c$ pair.   
In the two-quark process, the momentum transfer $\vec{q}$ is shared by
the heavy quark and the diquark in the heavy baryon, which may
excite both $\lambda$ and $\rho$ modes. 
This contrasts with the one-quark process where only one quark
receives the momentum transfer and therefore possible excitations
occurs only in the $\lambda$ modes.  

\begin{table}[htp]
\centering
\caption{Baryon masses $M$ in units of MeV, the spin-isospin
  coefficients for the heavy baryons $C_Y$, the relative magnitudes of
  the differential cross sections $\mathcal{R}(Y)$ that are normalized
  by that of the ground state $\Lambda(1/2^+)$. $Y_s$ and $Y_c$ denote
  the strange and charmed baryons, respectively. $j$ stands for the
  brown muck spin.}
\begin{tabular}{lccccccc}\hline\hline \\
$l=0$    & $\Lambda\left(\frac{1}{2}^+\right)$
     & $\Sigma\left(\frac{1}{2}^+\right)$
     & $\Sigma\left(\frac{3}{2}^+\right)$ & & & &  \\ 
\hline \\$M$[MeV] & 1116 & 1193 & 1385 &   &    &  & \\
       & 2286 & 2453 & 2518 &   &    &  & \\
$|C_Y|^2$&  1   & 3    & 0    &   &    &  & \\
$\mathcal{R}$($Y_s$) &  1   & 3.2  & 0    &   &    &  & \\
$\mathcal{R}$($Y_c$) &  1   & 2.9  & 0    &   &    &  & \\
\hline\hline \\
$l_\lambda=1$
& $\Lambda\left(\frac{1}{2}^-\right)$ 
& $\Lambda\left(\frac{3}{2}^-\right)$ 
& $\Sigma\left(\frac{1}{2}^-\right)$
& $\Sigma\left(\frac{1}{2}^-\right)$ 
& $\Sigma\left(\frac{3}{2}^-\right)$
& $\Sigma\left(\frac{3}{2}^-\right)$ 
& $\Sigma\left(\frac{5}{2}^-\right)$  \\
& $j=1$ & $j=1$ & $j=0$ & $j=1$ & $j=1$ & $j=2$ & $j=2$ \\
  \hline \\
  $M$[MeV]   & 1405 & 1520 & 1654 & 1734 & 1670 & 1755 & 1775 \\
         & 2595 & 2628 & 2802 & 2826 & 2807 & 2837 & 2839 \\
$|C_Y|^2$&  1/3 & 2/3  & 1/3  & 2/3  & 1/3  & 5/3  & 0\\
$\mathcal{R}$($Y_s$) & 0.0042& 0.0096& 0.0069& 0.015& 0.0070& 0.038 & 0\\
$\mathcal{R}$($Y_c$) & 0.10 & 0.20  & 0.12 & 0.23  & 0.12 & 0.58  & 0\\
\hline \hline\\
$l_\rho=1$  
& $\Lambda\left(\frac{1}{2}^-\right)$ 
& $\Lambda\left(\frac{1}{2}^-\right)$ 
& $\Lambda\left(\frac{3}{2}^-\right)$
& $\Lambda\left(\frac{3}{2}^-\right)$ 
& $\Lambda\left(\frac{5}{2}^-\right)$
& $\Sigma \left(\frac{1}{2}^-\right)$ 
& $\Sigma \left(\frac{3}{2}^-\right)$ \\
         & $j=0$ & $j=1$ & $j=1$ & $j=2$ & $j=2$ & $j=1$ & $j=1$ \\
  \hline\\
$M$[MeV]   & 1670 & 1777 & 1690 & 1810 & 1814 & 1751 & 1760 \\
       & 2890 & 2933 & 2917 & 2956 & 2960 & 2909 & 2910 \\
$|C_Y|^2$& 1/3  & 2/3  & 1/3  & 5/3  & 0  & 1/3  & 2/3  \\
$\mathcal{R}$($Y_s$) & 0.017& 0.039 & 0.018& 0.10 & 0   & 0.016& 0.032\\
$\mathcal{R}$($Y_c$) & 0.22 & 0.43  & 0.22 & 1.1  & 0   & 0.20 & 0.41  \\
\hline \hline
\end{tabular}
\label{tab:1}
\end{table}

We need numerical values of baryon masses with proper assignment of
the corresponding states to compute the cross sections.  
Actually, baryon masses in the quark model do not always agree with
the experimental data. For example, the mass of $\Lambda(1405)$ can
not be easily described by the quark model. So, we take the masses of
baryons from the Particle Data Group when
available~\cite{Patrignani:2016xqp}. Otherwise,  they are taken from
the values of the quark models~\cite{Yoshida:2015tia}. By using these
masses, we compute various matrix elements for the transitions up to
$p$-wave excitations. Results are shown in Table~\ref{tab:1}, where we
also list the masses of excited states, spin-isospin factors $|C_Y|^2$
and relative magnitudes of differential cross sections
$\mathcal{R}(Y)$ defined in Eq.~\eqref{eq:ProdRates}, which are
normalized by that of the ground-state $\Lambda(1/2^+)$. $Y_s$ and
$Y_c$ denote the strange and charmed baryons, respectively. $j$ stands
for the brown muck spin, which is the sum of the intrinsic spin and
the orbital angular momentum of a diquark. In the following
subsections, we will discuss the results in Table~\ref{tab:1} one by
one.  

\subsection{Production rates of ground and excited states}
We first discuss the difference between the production rates
of the strange and charmed baryons. In Table~\ref{tab:1}, we list the
results of the production rates for both the strange and charmed
baryons. As shown clearly in Table~\ref{tab:1}, the ground strange
baryons are more produced than the excited ones, whereas the
production rates of the excited charmed baryons are comparable with
those of the ground ones. In Ref.~\cite{Kim:2014qha} we see a similar
tendency. This can be understood by the dependence of the transition
amplitudes on the momentum transfer. Using the wavefunctions in the
basis of the harmonic oscillator, we are able to derive the matrix
elements analytically with Gaussian form factors depending on
$q^2_{\mathrm{eff}}$, which are given in   
Eq. (\ref{eq:TransAmpGn}), (\ref{eq:TransAmpLambda}) and
(\ref{eq:TransAmpRho}). The momentum transfer
$|\vec{q}_{\mathrm{eff}}|$ is given as a function of the initial and
final momenta, which depends on the total mass of the hadrons in the
final states. The squared \emph{effective} momentum transfer
$q^2_{\mathrm{eff}}$ governs the productions of the heavy baryons.
For example, the production rates of the lowest-lying heavy baryons
decrease as $q_{\mathrm{eff}}^2$ increases. It implies that in the
case of the productions of the ground-state heavy baryons, the
Gaussian form factor, $e^{-q_{\mathrm{eff}}^2/(4B^2)}$ mainly governs
the production mechanism.  On the other hand, when it comes to the
production rates of the excited states, $q_{\mathrm{eff}}^2$
dependence is much different from the case of the ground-state heavy
baryons.  In addition to the Gaussian form factor, there exist other
factors that are proportional to the $l$-th power of
$|\vec{q}_{\mathrm{eff}}|$, where $l$ denotes the orbital angular
momentum of the baryon in the final state. Thus, both the production 
rates for the $\rho$ and $\lambda$ modes are enhanced up 
to the maximum point as $q_{\mathrm{eff}}^2$ increases and then start
to fall off as $q_{\mathrm{eff}}^2$ further increases.

\begin{figure}[htp]
\captionsetup[subfigure]{labelformat=empty}
\subfloat[]{\includegraphics[width = 8cm]{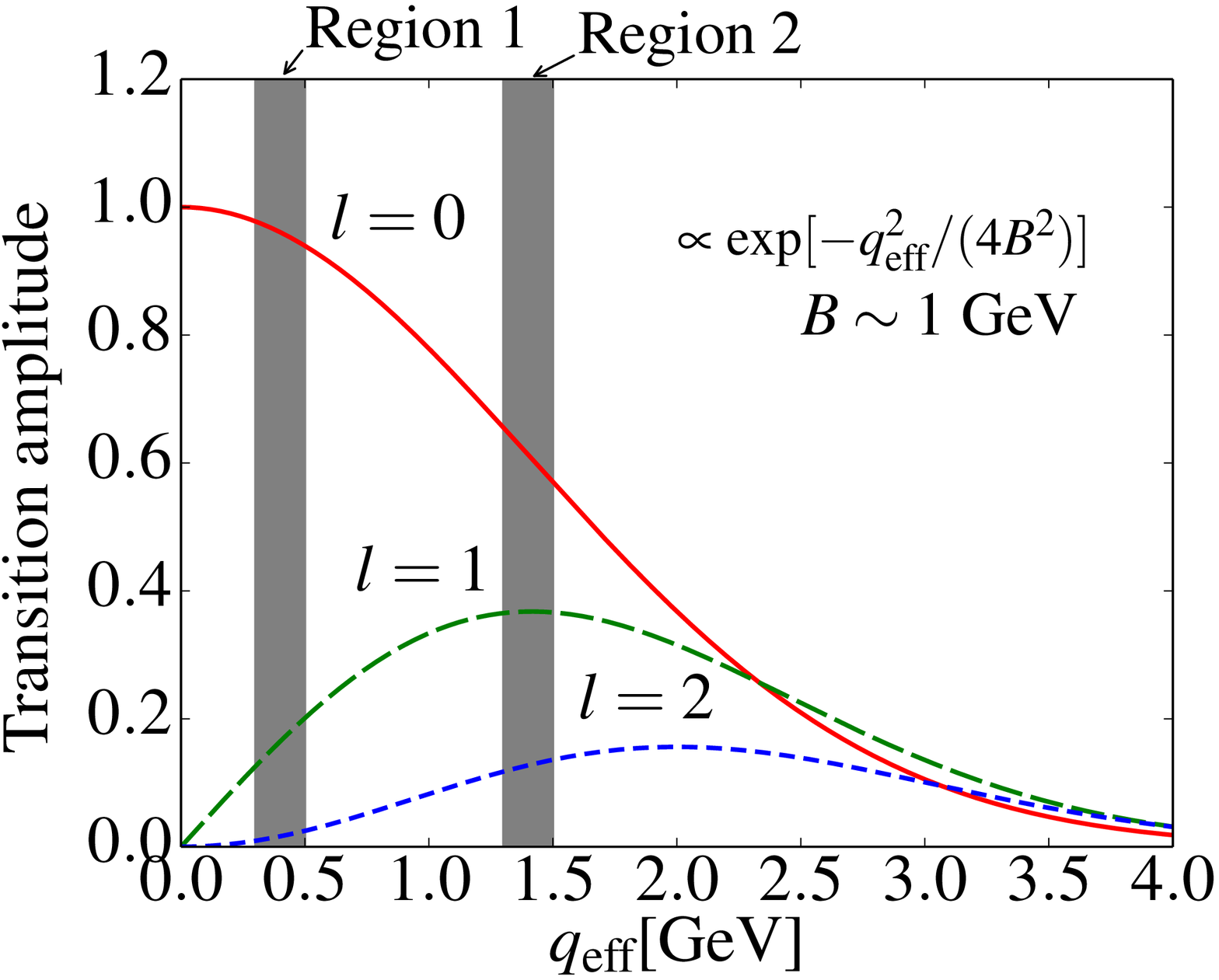}} 
\subfloat[]{\includegraphics[width = 8cm]{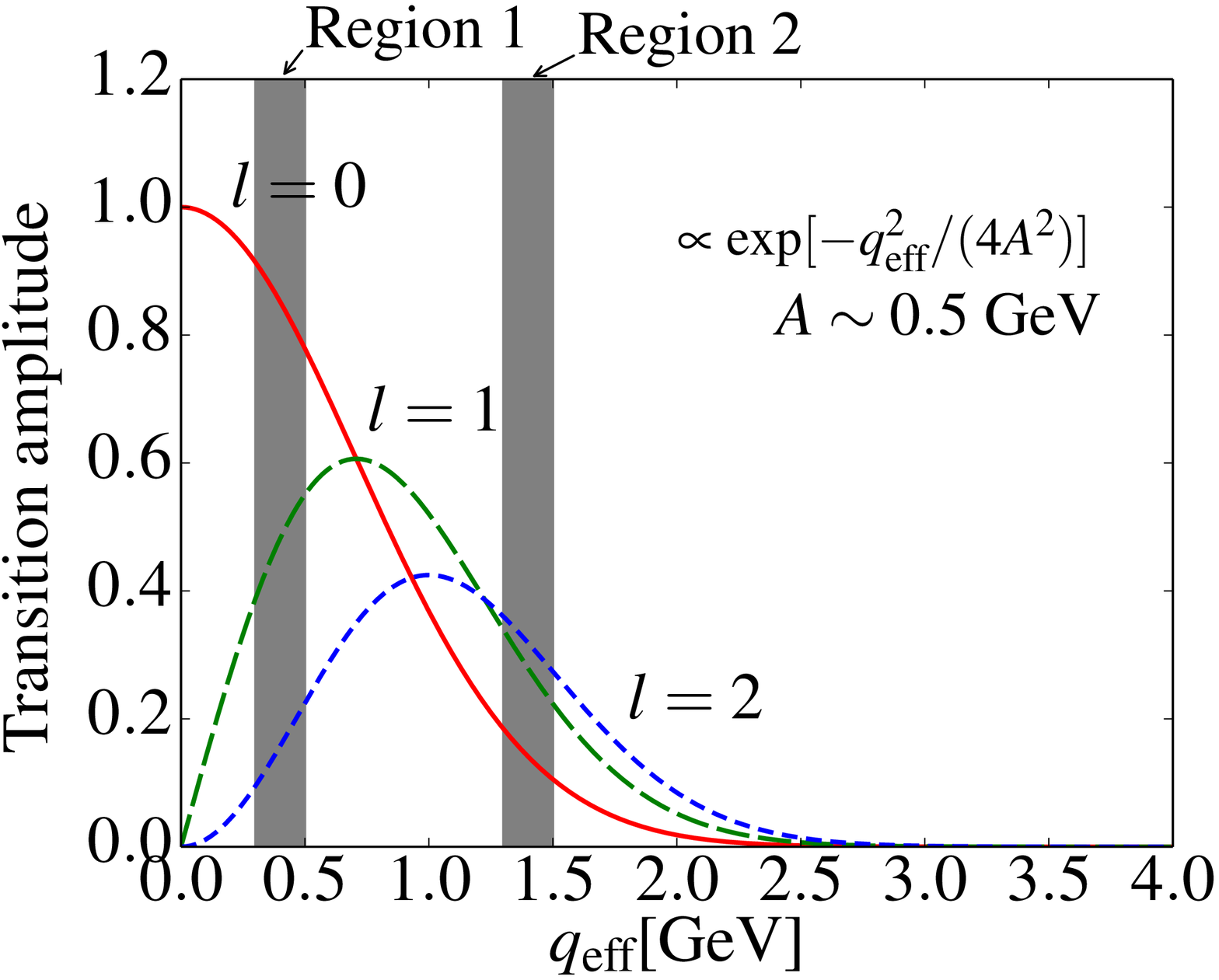}}
\caption{$|\vec{q}_{\mathrm{eff}}|$ dependences of the transition
    amplitudes with two-quark and the one-quark processes. The left
    panel is for the effects of the two-quark process with
    $B\simeq1$ GeV, whereas the right panel is for the contributions
    of the one-quark prosess with $A\simeq0.5$
    GeV. The solid curves, the long-dashed ones, and the short-dashed
    ones represent the contributions to the ground state ($l=0$), the
    $P$-wave and $D$-wave excited state, respectively. The gray shaded
    areas show the regions of the typical momentum transfers for
    strange and charmed baryon productions, Region 1 and Region 2, 
    respectively.}
\label{fig:4}
\end{figure}

To understand this feature more explicitly, let us examine 
various transition amplitudes as functions of the momentum transfer
$|\vec{q}_{\mathrm{eff}}|$. In the left panel of Fig.~\ref{fig:4}, we
show the normalized amplitudes for the transitions to $l=0$ (ground
state) and $1,\,2$ ($\lambda$ modes) baryons as functions of
$|\vec{q}_{\mathrm{eff}}|$ with Clebsh-Gordan coefficients
removed~\footnote{These definitions are different from those 
  in Ref.~\cite{Kim:2014qha} by $\sqrt{2}$ and $A$ is also replaced by
  $B$.},
  \begin{align}
I_0 &= e^{-q_{\mathrm{eff}}^2/(4 B^2)},\label{eq:I0} \\
I_1 &= \frac{1}{\sqrt{2}}
\Big(\frac{\alpha_{\lambda'}}{B}\Big)
|\vec{q}_{\mathrm{eff}}/B|e^{-q_{\mathrm{eff}}^2/(4 B^2)},\label{eq:I1} \\
I_2 &= \frac{1}{2\sqrt{3}}
\Big(\frac{\alpha_{\lambda'}}{B}\Big)^2
|\vec{q}_{\mathrm{eff}}/B|^2 e^{-q_{\mathrm{eff}}^2/(4 B^2)}.\label{eq:I2}
\end{align}
 For the strangeness production, the typical momentum transfer is shown by 
the \textit{Region 1}, 
where the ground state is the most abundantly produced, 
while for the charm production, as the \textit{Region 2} shows that 
the production rates of excited states become closer to that of the
ground state. 
\subsection{Two- vs. one-quark processes}
Here we briefly discuss the difference in the momentum dependences of 
transition amplitudes in the two-quark and one-quark processes.  
The amplitudes corresponding to  Eq.~(\ref{eq:I0})-(\ref{eq:I2}) for
the one-quark process is obtained by replacing the 
parameter $B$ by $A$, where
$A^2=(\alpha_{\lambda'}^2+\alpha_{\lambda}^2)/2$~\cite{Kim:2014qha}.   
Because of the exponential form factor $\exp(-q^2_{\rm{eff}}/(4
A^2\text{ or }4 B^2))$, the relation $B \sim 2 A$ implies that when
the momentum transfer 
becomes large the two-quark process dominates over the one-quark
process. Physically this is explained by the fact that the momentum
transfer is shared by two quarks rather than by one quark. By
comparing the two panels of Fig.~\ref{fig:4}, where the right panel is
for the result of the one quark process, this feature is
observed. For large momentum transfer $q_{\rm{eff}}>$ 2.5 GeV,
  transition amplitudes becomes negligibly small in the right panel
  while they are still considerable in the left one. So, the 
two-quark process is dominant over the one-quark process as
$q_{\mathrm{eff}}$ increases.

As listed in Table~\ref{tab:1} and shown in 
Fig. \ref{fig:4}, the production rates of various excited states of
the two-quark process are not as large as of those of the one-quark
process~\cite{Kim:2014qha}. A reason is in that the transition
amplitudes for the two-quark process are 
more broadly distributed to both the $\lambda$ and $\rho$ modes, while
the one-quark process contributes mainly to the $\lambda$ modes. 

\subsection{Transitions to $\lambda$ and $\rho$ modes of $\Lambda$
  and $\Sigma$ baryons}   
\begin{table}[htp]
\centering
\caption{The relative magnitudes of the differential cross sections
  for the $l=1$ excited states of the strange and charmed baryons,
  $\mathcal{R}(Y_s)$ and $\mathcal{R}(Y_c)$. $s$ and $j$ represent the
  intrinsic spin and the brown muck spin of the diquark. } 
\begin{tabular}{l|cc|ccccc}\hline\hline \\
$l_\lambda=1$
& $\Lambda\left(\frac{1}{2}^-\right)$ 
& $\Lambda\left(\frac{3}{2}^-\right)$ 
& $\Sigma\left(\frac{1}{2}^-\right)$
& $\Sigma\left(\frac{1}{2}^-\right)$ 
& $\Sigma\left(\frac{3}{2}^-\right)$
& $\Sigma\left(\frac{3}{2}^-\right)$ 
& $\Sigma\left(\frac{5}{2}^-\right)$  \\
& $s=0$ & $s=0$ & $s=1$ & $s=1$ & $s=1$ & $s=1$ & $s=1$ \\
& $j=1$ & $j=1$ & $j=0$ & $j=1$ & $j=1$ & $j=2$ & $j=2$ \\
$\mathcal{R}$($Y_s$) & 0.0042& 0.0096& 0.0069& 0.015& 0.0070& 0.038 & 0\\
$\mathcal{R}$($Y_c$) & 0.10 & 0.20  & 0.12 & 0.23  & 0.12 & 0.58  & 0\\
$Ratio$ & 1 & 2  & 1 & 2  & 1 & 5  & 0  \\
\hline \hline\\
$l_\rho=1$  
& $\Sigma \left(\frac{1}{2}^-\right)$ 
& $\Sigma \left(\frac{3}{2}^-\right)$
& $\Lambda\left(\frac{1}{2}^-\right)$ 
& $\Lambda\left(\frac{1}{2}^-\right)$ 
& $\Lambda\left(\frac{3}{2}^-\right)$
& $\Lambda\left(\frac{3}{2}^-\right)$ 
& $\Lambda\left(\frac{5}{2}^-\right)$ \\
& $s=0$ & $s=0$ & $s=1$ & $s=1$ & $s=1$ & $s=1$ & $s=1$ \\
& $j=1$ & $j=1$ & $j=0$ & $j=1$ & $j=1$ & $j=2$ & $j=2$ \\
$\mathcal{R}$($Y_s$) & 0.016& 0.032 & 0.017& 0.039 & 0.018& 0.10 & 0 \\
$\mathcal{R}$($Y_c$) & 0.20 & 0.41  & 0.22 & 0.43  & 0.22 & 1.1  & 0  \\
$Ratio$ & 1 & 2  & 1 & 2  & 1 & 5  & 0  \\
\hline \hline
\end{tabular}

\label{tab:2}
\end{table}
In order to discuss the relations between production rates
and the spin structures, we want to examine the production
rates of $\lambda$ and $\rho$ modes of $\Lambda$ and $\Sigma$
baryons. Table~\ref{tab:2} reorganizes relevant differential cross
sections $\mathcal{R}(Y)$ taken from Table~\ref{tab:1} and roughly
estimated ratios in each group. Here, $s$ and 
$j$ denote respectively the spin of the light diquarks and the spin of
the brown muck, which are just the coupled angular momentum of the
diquark spin and its orbital angular momentum. If we scrutinize the
results listed in Table~\ref{tab:2}, we can observe a systematics in
$\lambda$- and $\rho$-mode productions. Namely, the ratio of the
$\Lambda$ baryons of the $\lambda$ modes is $ 1 : 2$ and it is same as
that of $\Sigma$ baryons of the $\rho$ modes, and that of the
$\Lambda$ baryons of the $\rho$ modes which is $1 : 2 : 1 : 5 : 0$
coincides with that of $\Sigma$ baryons of the $\lambda$ modes.
Considering the values of $s$ and $j$, we find that the excited
$\Lambda$ baryons in the $\lambda$ mode have the similar spin
  structures which have same quantum numbers, $s$, $j$, and $J^p$, to   
those of the excited $\Sigma$ baryons in the $\rho$ mode. Similarly,
the excited $\Sigma$ baryons in the $\lambda$ mode correspond to the
excited $\Lambda$ baryons in the $\rho$ mode by the spin
content. The explicit forms of the 
wave functions can be found by using Eqs.~\eqref{e:WFLambda} and
\eqref{e:WFSigma}. Thus, the identity of a baryon either in the
$\lambda$ mode or in the $\rho$ mode is determined by the study of
production rates.    

\subsection{Restriction on the spin due to the instanton interaction}  
We want to mention that in the present work the spin flip of the quark
does not occur during the process of the baryon productions, because
the leading terms in the $1/N_c$ expansion of the 't Hooft-like
interaction are spin independent. This restricts the transition
processes by certain conditions. As already shown in
Table~\ref{tab:1}, the excited hyperons $\Sigma(\frac{3}{2}^+)$,
$\Sigma(\frac{5}{2}^-)$ and 
$\Lambda(\frac{5}{2}^-)$ are not allowed to be produced off the
proton.  The absence of spin-flip interactions keeps the intrinsic
spins of the quarks intact, which implies that the excitations of the
orbital angular momenta cannot produce the above-mentioned
excited hyperons. The intrinsic spins of the quarks inside a proton
can be flipped only by the vector or tensor interactions in the course
of the production processes. Thus, we need to consider the vector or
tensor interactions that make the intrinsic spins flipped. We will
leave it as a future work.  

\subsection{Production rates of $\Lambda$'s and $\Sigma$'s}  
There is yet another interesting point in the present results: 
we find that the ground-state $\Sigma$ baryons are in general produced
more abundantly than the corresponding $\Lambda$ ones. 
As shown in Table~\ref{tab:1}, we have obtained the ratio of 
$\Lambda(\frac{1}{2}^+)$ to $\Sigma(\frac{1}{2}^+)$ is around
$1/3$, while the previous study~\cite{Kim:2014qha}, in which the one-quark
process was only taken into account for the productions of the heavy
baryons and vector mesons, yielded the results opposite to 
the present one, i.e. the corresponding ratio turns out around $30$. 

These ratios reflect the spin and isospin structures of the reaction 
mechanism due to the relevant operators and wave functions.   
In this regard, it is interesting to observe that the ratio $1/3$
holds also for the transitions to excited states; the sums of the
transitions to the $\lambda$ modes of $\Lambda$'s and $\Sigma$'s, and
those of the $\rho$ modes of the $\Sigma$'s and $\Lambda$'s. 
Note that the available experimental data show that the ratio between
the $\Lambda(\frac{1}{2}^+)$ and $\Sigma(\frac{1}{2}^+)$ productions
is given around $3/2$~\cite{Crennell:1972km}.  It implies that both
the one-quark and two-quark processes should be taken into account to 
describe the existing data of $\Lambda\left(\frac{1}{2}^+\right)$ and
$\Sigma\left(\frac{1}{2}^+\right)$. 
The relative strength of one-quark and two-quark processes may be
determined by an additional study of the one-quark process for the
productions of heavy baryons and pseudoscalar mesons or it is also
possible by that of the two-quark process for heavy baryons and vector
mesons with the previous study~\cite{Kim:2014qha} as well.  
It will be possible to carry out
more detailed studies, when features of different reaction mechanisms
will be understood better.

\section{Summary and conclusions}
In the present work, we aim at investigating the productions of
strange and charmed baryons, including both the one-quark and
two-quark processes. While the one-quark process was already
considered previously, the two-quark process was proposed in this
work. By the two-quark process, we mean that the two quarks inside a
baryon undergo the interaction with a quark inside a meson beam, so
that a strange or charmed baryon is produced. Thus, we need to
introduce the three-quark interaction involving both the light and
heavy quarks. In order to realize this three-quark interaction, we
introduced a 't Hooft-like interaction arising from the instanton
vacuum. The six-quark operators in the 't Hooft-like interaction were
decomposed into the quark fields for the mesons and those for the
baryons. To make the investigation simpler, we construct the baryon
wave functions based on the nonrelativistic quark model with the
confining potential of the harmonic-oscillator type. 
The excitations of the produced baryons consist  
of the two modes, i.e. the $\lambda$ mode and
the $\rho$ mode. As already shown in previous works, the one-quark
process excites only the $\lambda$ mode. However, the two-quark
process does both the $\lambda$ and $\rho$ modes. Thus, the two-quark
process allows one to scrutinize the production mechanism of the
excited charmed baryons in a more microscopic way. In particular, when
the momentum transfer becomes large, the two-quark process will come 
into more important play. However, since introducing three-quark
interactions involve additional ambiguity from unknown parameters, we
mainly focussed on the ratios of the production cross sections between
the strange and charmed baryons in the present work.  

The main results are summarized as follows:
\begin{itemize}
\item The excited states are more produced for the charmed baryons  than
for the strange baryons (hyperons), which was also found in the
previous work. This can be understood by examining the dependence of
the transition amplitudes on the momentum transfer. The amplitudes
show the additional dependence on the momentum transfer, which arises
from the higher orbital angular momentum. 

\item The two-quark processes excite not only the $\lambda$ modes but
  also the $\rho$ modes, which is distinguished from the one-quark
  processes. 

\item The production rates reflect the spin structure of baryons.  
For instance, the relative production rates of $\lambda$-mode
$\Lambda$'s are similar to those of $\rho$-mode $\Sigma$'s, because 
they have similar spin structures. These relations can be used for
the identification of newly found baryons with unknown spin structure.

\item For the ground-state heavy baryons, $\Sigma$'s are more produced
  than $\Lambda$'s. The one-quark processes exaggerate the relative
  production rates of the $\Sigma$'s in comparison with
  $\Lambda$'s, since the observed ground-state $\Sigma$ production
  rates are about half of those of the $\Lambda$ hyperons. It implies
  that both the one-quark and two-quark processes come into play to
  describe the production mechanism of the hyperons. Thus, the
  two-quark processes should be considered as much as the
  one-quark processes. 
\end{itemize}

In the present work, we study the productions of the strange and
charmed baryons in a qualitative manner. To investigate the production
mechanisms of those baryons, we have to investigate the following
issues.    
\begin{itemize}
\item The instanton-induced interactions provide scalar-type
  interaction in the leading order of $1/N_c$ expansion. However, the
  inclusion of the $1/N_c$ corrections is inevitable to describe the
  spin-flipped processes. Moreover, it is of great importance to
  introduce vector or tensor interactions for the baryon production in
  high-energy processes, as the Regge theories already implied. 

\item The present study was mainly focussed on the forward angle
  productions. We need to cover the whole angle to investigate the
  productions of strange and charmed baryons in a more quantitative
  way. 

\item The study of the baryon productions aim eventually at extracting
  information the structures of the baryons concerned. Thus, it is of
  great interest to implement microscopically the effects of the
  diquark and multi-quark structure in the description of the baryon
  productions. 
\end{itemize}
All these issues mentioned above will be discussed in forthcoming
works. 

\section*{Acknowledgments}
We thank H. Noumi and K. Shirotori for useful discussions about
experimental situations.   
This work is supported in part by Grants-in-Aid for Scientific
Research Grants No. JP17K05441 (C) and by Scientific Research on
Innovative Areas No. 18H05407.
The work of S.-I. S is supported by Rotary Yoneyama Memorial Foundation.
The work of H.-Ch. K is supported by
Basic Science Research Program through the National Research
Foundation (NRF) of Korea funded by the Ministry of Education, Science and
Technology (MIST) (No. 2018R1A5A1025563 and No. NRF-2018R1A2B2001752).

\appendix
\section{Radial part baryon wave functions}
The radial part of baryon wave fuctions $R_{n l}(r)$ are given with
the wave functions of 3D harmonic oscillators as following, 
\begin{align}
R_{0 0}(r)
  &=\left(\frac{4\alpha^{3}}{\sqrt{\pi}}\right)^{1/2}e^{-(\alpha
    r)^2/2}, \cr
R_{0 1}(r) &= \left(\frac{8\alpha^{3}}{3\sqrt{\pi}}\right)^{1/2}\alpha
             r e^{-(\alpha r)^2/2},  
\end{align}
where the oscillator parametor $\alpha$ is given as
\begin{align}
\alpha_{\rho} &= \left( \frac{3 k}{4 m_q} \right)^{1/4}
\end{align}
for the $\rho$-mode wave functions of baryons and
\begin{align}
\alpha_{\lambda} &= \left( \frac{4 k}{3 m_q} \right)^{1/4}, \nonumber \\
\alpha_{\lambda'} &= \left( \frac{2 (m_d + m_Q) k}{m_d m_Q} \right)^{1/4}
\end{align}
for the $\lambda$-mode wave functions of the inital and final state
baryons, respectively. Here, $k$ is the spring constant between quarks.
\section{Integrations with Gaussian integrals}
To find the final expressions of Eq.(\ref{eq:Igs}),
(\ref{eq:Ilambda1}) and (\ref{eq:Ilambda1}), we use the Gaussian
integrals. Some parts of the derivations for $I_{g.s.}$ and
$I_{\lambda=1}$ are given as following. 
\begin{align}
I_{g.s.} &\equiv
\int d^3 \kappa \,
\int d^3 \rho\, e^{i \frac{1}{2}\vec{\kappa} \cdot \vec{\rho}}
\psi^{\rho *}_{0}(\vec{\rho}) \psi^{\rho }_{0} (\vec{\rho})
\int d^3 \lambda\, e^{i (\vec{\kappa}+\vec{q}_{\mathrm{eff}}) \cdot \vec{\lambda}}
\psi^{\lambda' *}_{0}(\vec{\lambda}) \psi^{\lambda}_{0}
           (\vec{\lambda})\nonumber\\ 
&=
\int d^3 \kappa \,
\left(
\frac{\alpha_{\rho}^2}{\pi}
\right)^{3/2}
\int d^3 \rho\, \exp 
\left[
-\alpha^2_{\rho} \rho^2 + i \frac{1}{2}\vec{\kappa} \cdot \vec{\rho}
\right]
\nonumber\\
&\hspace{0.4cm}\times
\left(
\frac{ \alpha_{\lambda} \alpha'_{\lambda}}{\pi}
\right)^{3/2}
\int d^3 \lambda\,
\exp 
\Big[
-\alpha^2_{\lambda} \lambda^2 
+ i (\vec{\kappa} +\vec{q}_{\mathrm{eff}}) \cdot \vec{\lambda}
\Big]
\nonumber\\
&=
\int d^3 \kappa \,
\exp
\Big[
-\frac{\kappa^2}{16\alpha^2_{\rho}}
\Big]
\left(
\frac{2\alpha_{\lambda} \alpha'_{\lambda}}{
\alpha_{\lambda} + \alpha'_{\lambda}}
\right)^{3/2}
\exp 
\left[
-
\frac{(\vec{\kappa} +\vec{q}_{\mathrm{eff}})^2}{2(\alpha^2_{\lambda} +
  \alpha^{'2}_{\lambda})} 
\right]
\nonumber\\
&=
\left(
\frac{2\alpha_{\lambda} \alpha'_{\lambda} }{\alpha_{\lambda} + \alpha'_{\lambda}}
\right)^{3/2}
\exp 
\left[
-
\frac{q^2_{\mathrm{eff}}}{2(\alpha^2_{\lambda} + \alpha^{'2}_{\lambda})}
\right]
\int d^3 \kappa \,
\exp
\Big[
-\frac{B^2}{
8\alpha^2_{\rho}(\alpha^2_{\lambda}+\alpha^{'2}_{\lambda})}\kappa^2
-\frac{\vec{q}_{\mathrm{eff}}}{
\alpha^2_{\lambda}+\alpha^{'2}_{\lambda}} \cdot \vec{\kappa}
\Big]
\nonumber\\
&=
\left(
\frac{16\pi \alpha^2_{\rho} \alpha_{\lambda'}\alpha_{\lambda}}{
	B^2}
\right)^{3/2}
e^{-q^2_{\mathrm{eff}}/(4 B^2)},\\
I_{l_\lambda=1} &\equiv 
\int d^3 \kappa \,
\int d^3 \rho e^{i \frac{1}{2}\vec{\kappa} \cdot \vec{\rho}}
\psi^{\rho *}_{0}(\vec{\rho}) \psi^{\rho }_{0} (\vec{\rho})
\int d^3 \lambda e^{i (\vec{\kappa}+\vec{q}_{\mathrm{eff}}) \cdot \vec{\lambda}}
\psi^{\lambda' *}_{1}(\vec{\lambda}) \psi^{\lambda}_{0} (\vec{\lambda})\nonumber\\
&=
\int d^3 \kappa \,
\exp
\Big[
-\frac{\kappa^2}{16\alpha^2_{\rho}}
\Big]
\sqrt{2}\alpha'_{\lambda}
\left(
\frac{ \alpha_{\lambda} \alpha'_{\lambda}}{\pi}
\right)^{3/2}
\int d^3 \lambda\, 
\lambda_z \exp
\Big[
-\alpha^2_{\lambda} \lambda^2 
+ i (\vec{\kappa} +\vec{q}_{\mathrm{eff}}) \cdot \vec{\lambda}
\Big]
\nonumber\\
&=
\int d^3 \kappa \,
\exp
\Big[
-\frac{\kappa^2}{16\alpha^2_{\rho}}
\Big]
\sqrt{2}\alpha'_{\lambda}
\left(
\frac{ 2 \alpha_{\lambda} \alpha'_{\lambda}}{
\alpha_{\lambda}^2 + \alpha'^2_{\lambda}}
\right)^{3/2}
\left(
i\frac{ 
\kappa_{ z} + (q_{\mathrm{eff}})_z
}{
\alpha_{\lambda}^2 + \alpha'^2_{\lambda}}
\right)
\exp
\left[
-\frac{
	(\vec{\kappa}+\vec{q}_{\mathrm{eff}})^2
}{
	2 (\alpha_{\lambda}^2+\alpha_{\lambda}^2)
}
\right]
\nonumber\\
&=
\frac{
	i\sqrt{2}\alpha'_{\lambda}
}{
	\alpha_{\lambda}^2+\alpha_{\lambda}^2	
}
\left(
\frac{ 2 \alpha_{\lambda} \alpha'_{\lambda}}{
\alpha_{\lambda}^2 + \alpha'^2_{\lambda}}
\right)^{3/2}
\exp 
\left[
-
\frac{q^2_{\mathrm{eff}}}{2(\alpha^2_{\lambda} + \alpha^{'2}_{\lambda})}
\right]
\nonumber\\
&\hspace{0.4cm}\times
\int d^3 \kappa \,
\Big( \kappa_{z} + (q_{\mathrm{eff}})_z \Big)
\exp
\Big[
-\frac{B^2}{
8\alpha^2_{\rho}(\alpha^2_{\lambda}+\alpha^{'2}_{\lambda})}\kappa^2
-\frac{\vec{q}_{\mathrm{eff}}}{
\alpha^2_{\lambda}+\alpha^{'2}_{\lambda}} \cdot \vec{\kappa}
\Big]
\nonumber\\
&=
\frac{i\sqrt{2}\alpha_{\lambda'} (q_{\mathrm{eff}})_z}{
	2 B^2}
\left(
\frac{16\pi \alpha^2_{\rho} \alpha_{\lambda'}\alpha_{\lambda}}{
	B^2}
\right)^{3/2}
e^{-q^2_{\mathrm{eff}}/(4 B^2)}
\end{align}

Here, the following formulae of the integrals have been used for
integrating over $\rho$, $\lambda$ and $q_1$. 
\begin{align}
\int d r\, e^{-Ar^2+\vec{B}\cdot\vec{r}}
&=\left(
\frac{\pi}{A}
\right)^{3/2}
e^{\frac{B^2}{4A}} \\
\int d r\, r_i\, e^{-Ar^2+\vec{B}\cdot\vec{r}}
&=
\frac{B_i}{2A}
\left(
\frac{\pi}{A}
\right)^{3/2}
e^{\frac{B^2}{4A}}
\end{align}


\end{document}